\documentclass[twocolumn,showpacs,showkeys,amsmath,amssymb]{revtex4}

\usepackage{graphicx}
\usepackage{dcolumn}
\usepackage{bm}

\begin{document}

\title{Confined Ge--Pt states in self-organized Pt nanowire arrays on Ge(001)}

\author{U.~Schwingenschl\"ogl and C.~Schuster}
\affiliation{Institut f\"ur Physik, Universit\"at Augsburg, 86135 Augsburg, Germany}

\date{\today}


\begin{abstract}
By means of band structure calculations within the density functional
theory and the generalized gradient approximation, we investigate the
electronic structure of self-organized Pt nanowires on the Ge(001) surface.
In particular, we deal with a novel one-dimensional surface state confined
in the nanowire array and clarify its origin. Due to large Pt contributions,
the novel state is rather a mixed Ge--Pt hybrid state than a confined Ge
surface state. Moreover, we compare our results to data from scanning
tunneling microscopy.
\end{abstract}

\maketitle

Because of a wealth of extraordinary physical properties, self-organized
nanowires recently attract great attention. Of special importance are
self-organization phenomena on semiconductor surfaces, due to a wide
field of possible technological applications. For the Ge(001) surface,
for example, Au growth comes along with a large variety of ordering
processes as a function of the Au coverage and growth temperature \cite{wang04,wang05}.
Moreover, adsorption of Pt atoms results in well-ordered chain arrays
after high-temperature annealing \cite{gurlu03}. These spontaneously
formed Pt nanowires are promising with respect to their thermodynamical
stablity, regularity, and length, which reaches up
to several hundred nanometers. In order to obtain detailed insight into
their physical properties and the self-organization process,
exact knowledge about the surface electronic states is mandatory, which
we address in the present paper.

A one-dimensional electronic state on the Pt-covered Ge(001)
surface has been discovered by Oncel {\it et al.} \cite{oncel05}.
Using scanning tunneling microscopy, the authors have measured spatial maps
of the differential conductivity and afterwards calculated from these data
the local density of states (DOS). At a temperature of 77\,K and for a nanowire spacing
of 1.6\,nm, they find an unexpected DOS peak some $0.1$\,eV above the Fermi
level. They argue that this electronic state is located in the Ge trough
between the Pt nanowires and, therefore, attribute it to a confined Ge
surface state. However, the Ge(001) surface is known to be subject to
strong hybridization under Pt coverage \cite{GePt}. For this reason, we
subsequently investigate the electronic structure of the Pt-covered Ge(001) surface
by means of electronic structure calculations based on density functional
theory. The band structure results discussed in the following rely on a fully
relaxed supercell of the Ge--Pt surface, where the distance between
neighbouring Pt nanowires has been set to 1.6\,nm in order to prepare
for a comparison of our first principles findings to the experimental DOS.

\begin{figure}[t]
\includegraphics[width=0.45\textwidth,clip]{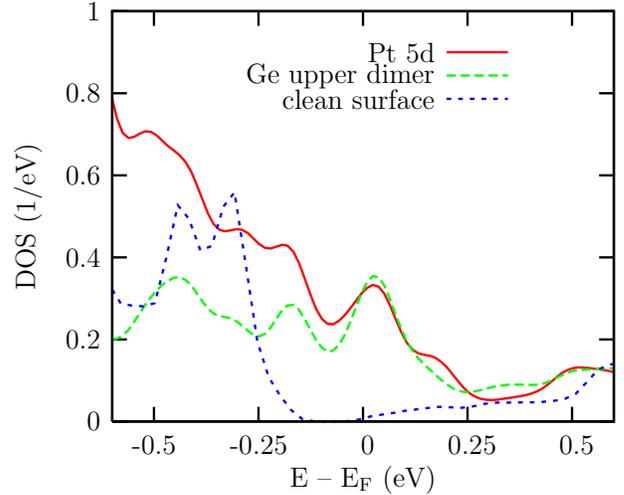}
\caption{(Color online) Partial Pt $5d$ and Ge densities of states (per atom) in the
vicinity of the Fermi energy. The Ge DOS refers to the upper dimer site and is shown
both for a clean Ge(001) surface and a surface covered with Pt nanowires. For convenience,
the Pt $5d$ DOS is downscaled by a factor 1/2.}
\label{fig1}
\end{figure}

We use the generalized gradient approximation (GGA) implemented in the
full-potential linearized augmented-plane-wave WIEN2k code
\cite{wien2k}, which is known for a very high capability in dealing with
surfaces/interfaces \cite{GePt,sur-int}. The exchange-correlation
potential here is parametrized according to the Perdew-Burke-Ernzerhof
scheme. Our data is based on a supercell of the cubic Ge unit cell
consisting of a $c(4\times2)$ reconstructed surface array and extending
two unit cells perdendicular to the surface. The formation of Pt chains
on Ge(001) is accompanied by the partial breakup of the Ge surface
dimers \cite{gurlu03}. For obtaining a useful starting point for
the structure optimization we thus start from the clean Ge surface,
leave away each fourth row of upper Ge dimer sites, and place Pt
atoms in the trough next to that row. Finally,
we assume convergence of the structure optimization when the surface
forces have decayed below a threshold of 5\,mRyd/$a_B$.

\begin{figure}[t]
\includegraphics[width=0.45\textwidth,clip]{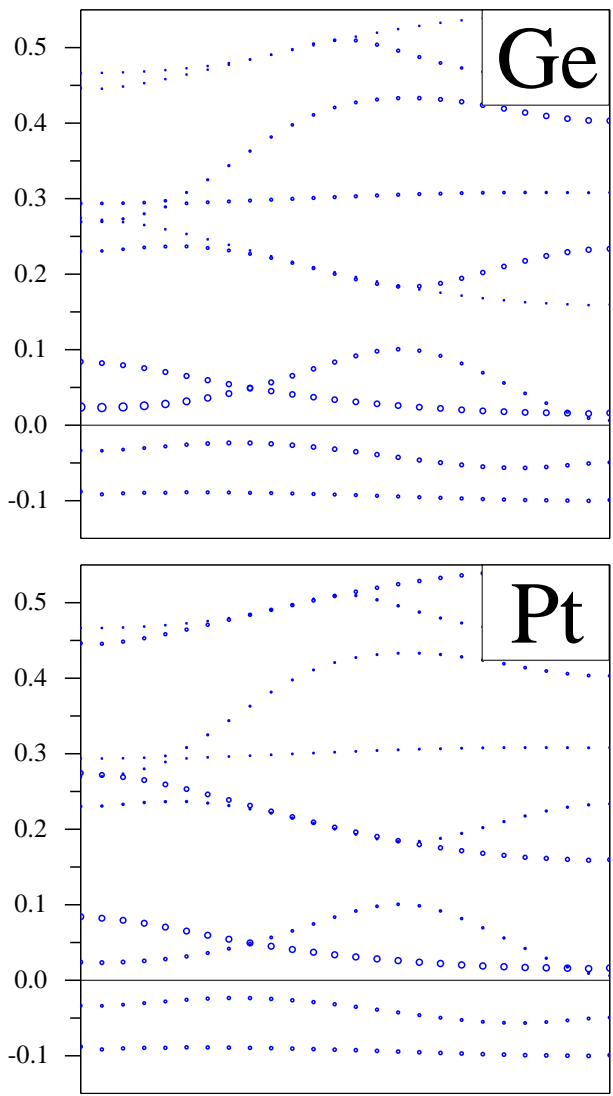}
\caption{(Color online) Weighted electronic bands highlighting
Ge and Pt $5d$ contributions via the size of the circles
drawn for each band and {\bf k}-point. Covering the first Brillouin zone,
the representation refers to the direction perpendicular to the Pt nanowires
in the surface plane.}
\label{fig2}
\end{figure}

Fig.\ \ref{fig1} shows partial Pt $5d$ and Ge densities of states as resulting from
our electronic structure calculation. As an example, we address the DOS of the upper
dimer Ge site. However, the following discussion does not depend on this choice, as
similar results are obtained for all Ge surface sites. For comparison, Fig.\ \ref{fig1}
includes the upper dimer Ge DOS for the clean Ge(001) surface without Pt nanowires
\cite{Ge}. Confirming the findings of Oncel {\it et al.} \cite[Fig.\ 2b]{oncel05},
for the Pt-covered surface a distinct structure appears in the Ge DOS just above
the Fermi level, which we therefore ascribe to the novel electronic state under
consideration. In the theoretical DOS, it is found at an energy of some $0.05$\,eV,
rather than at the experimental value of $0.1$\,eV. Amazingly, the $0.05$\,eV peak
reappears in the Pt $5d$ DOS with a very similar shape. While the Ge DOS in Fig.\
\ref{fig1} is normalized with respect to the number of atoms contributing, the Pt $5d$
DOS additionally has been downscaled by a factor 1/2. As a consequence, a significant
amount of the spectral weight in the vicinity of the Fermi energy has to be attributed to the Pt
nanowires. Furthermore, careful analysis of the band structure reveals that mainly two
electronic bands of mixed orbital character give rise to the DOS structure around
$0.05$\,eV. Such a strong hybridization clearly points at significant Ge--Pt interaction.
Fig.\ \ref{fig2} depicts the band structure in the relevant energy range,
where either the Ge or Pt $5d$ contributions to the states are highlighted by
means of the point size. The representation refers to the first Brillouin zone, and the
direction parallel to the surface plane and perpendicular to the nanowires.
A tiny band gap of approximately 0.03\,eV is observed in fig.\ \ref{fig2},
reflecting the fact that the conductivitiy of the Pt-covered Ge(001)
surface is basically connected to the nanowire formation.

\begin{figure}[t]
\includegraphics[width=0.45\textwidth,clip]{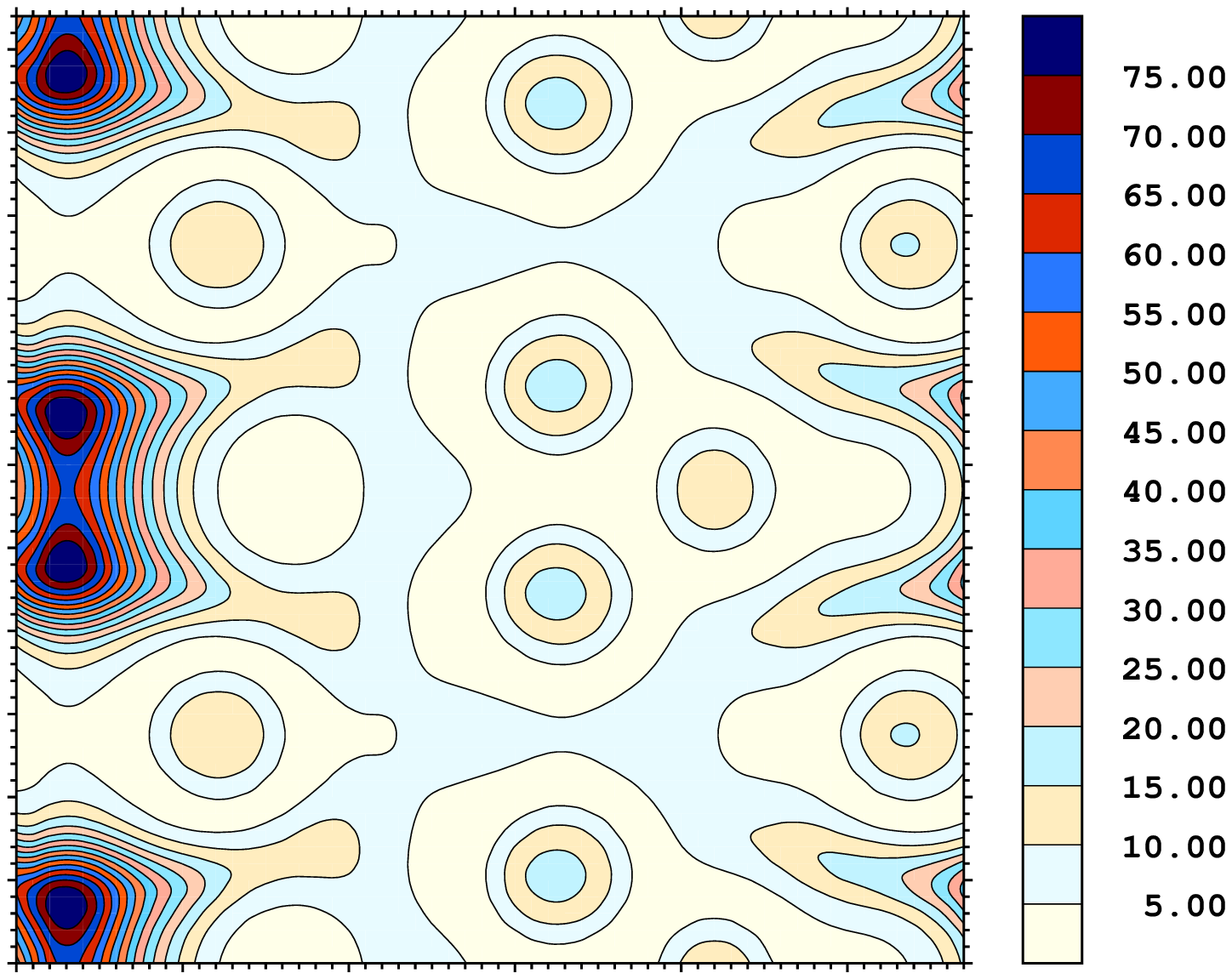}
\caption{(Color online) Simulated STM image,
based on the occupied DOS in the energy range from
$-1$\,eV to the Fermi level. The valence states of the Pt chains
(center of the figure, running from top to bottom) reveal a distinct
dimerization.}
\label{fig3}
\end{figure}

While the existence of a novel surface electronic state is confirmed,
our band structure data spoil an interpretation in terms of a confined Ge surface
state. Instead, as large Pt admixtures are due to strong Ge--Pt
hybridization, the state must originate from both the Ge surface and the
Pt nanowires. Supporting our line of reasoning, scanning tunneling
microscopy (STM) data, as obtained by Sch\"afer {\it et al.} \cite{schafer06}, show
that the Pt conduction bands are seriously modified by the interaction
with the Ge substrate. For comparison, we present a simulted STM
image in fig.\ \ref{fig3}. The image indicates a remarkable
dimerization of the Pt valence states, which are seen in the very center
of fig.\ \ref{fig3}, running from top to bottom. However, contradicting
the experiment, this dimerization does not vanish for states close to
the Fermi level. Modelling the surface energy levels in terms of a
quantum mechanical particle in a well, as proposed in \cite{oncel05},
does not suffer from our interpretation in terms of mixed
Ge--Pt states. Exactly the same applies to the fact that the
state gradually fades away close to structural defects of the nanowires
or the substrate. In contrast, the experimental observation
of an almost perfect localization within the Ge trough is puzzeling
because of the large Pt admixtures. However, the simple shape of the
band with the largest Ge--Pt hybridization in fig.\ \ref{fig2}
indicates that the novel surface state is captured by the classical
Kronig--Penney model. Thus, it is well described in terms of a
non-localized Ge--Pt state confined in the periodic potential of the nanowire array.

In conclusion, we have presented first principles band structure
calculations for the Pt-covered Ge(001) surface and studied
the surface electronic structure. In particular, we have
focussed on a novel electronic state, located right above the Fermi energy,
which is not found for the clean Ge(001) surface. Contradicting a
previous experimental prediction, our data show that this state does
not originate from simple confinement of a regular Ge surface state.
For an adequate interpretation of the experimental DOS it is essential
to account for strong Ge--Pt hybridization, which is not surprising
because interaction between the substrate and the adsorbate is
expected to be of great importance for the adsorption process \cite{GePt,schmitt05}.

\vspace{0.2cm}
Financial support by the Deutsche Forschungsgemeinschaft (SFB 484) is acknowledged.


\begin{thebibliography}{10}

\bibitem{wang04}
J.\ Wang, M.\ Li, and E.I.\ Altman,
Phys.\ Rev.\ B {\bf 70}, 233312 (2004).

\bibitem{wang05}
J.\ Wang, M.\ Li, and E.I.\ Altman,
Surf.\ Sci.\ {\bf 596}, 126 (2005).

\bibitem{gurlu03}
O.\ Gurlu, O.A.O.\ Adam, H.J.W.\ Zandvliet, and B.\ Poelsema,
Appl.\ Phys.\ Lett.\ {\bf 83}, 4610 (2003).

\bibitem{oncel05}
N.\ Oncel, A.\ van Houselt, J.\ Huijben, A.-S.\ Hallb\"ack, O.\ Gurlu,
H.J.W.\ Zandvliet, and B.\ Poelsema, Phys.\ Rev.\ Lett.\ {\bf 95}, 116801 (2005).

\bibitem{GePt}
U.\ Schwingenschl\"ogl and C.\ Schuster, Europhys.\ Lett.\ {\bf 81}, 26001 (2008).

\bibitem{sur-int}
U.\ Schwingenschl\"ogl and C.\ Schuster, Chem.\ Phys.\ Lett.\ {\bf 439},
143 (2007); Appl.\ Phys.\ Lett.\ {\bf 90}, 192502 (2007); Europhys.\ Lett.\ {\bf 81}, 17007 (2008).

\bibitem{wien2k}
P.\ Blaha, K.\ Schwarz, G.\ Madsen, D.\ Kvasicka, and J.\ Luitz, WIEN2k: An
augmented plane wave and local orbitals program for calculating crystal properties
(Vienna University of Technology, Austria, 2001).

\bibitem{Ge}
U.\ Schwingenschl\"ogl and C.\ Schuster, Chem.\ Phys.\ Lett.,
{\bf 449} 126 (2007).

\bibitem{schafer06}
J.\ Sch\"afer, D.\ Schrupp, M.\ Preisinger, and R.\ Claessen,
Phys.\ Rev.\ B {\bf 74}, 041404(R) (2006).

\bibitem{schmitt05} T.\ Schmitt, A.\ Augustsson, J.\ Nordgren, L.-C.\ Duda, J.\ H\"owing,
T.\ Gustafsson, U.\ Schwingenschl\"ogl, and V.\ Eyert,
Appl.\ Phys.\ Lett.\ {\bf 86}, 064101 (2005).

\end{thebibliography}
\end{document}